# Predominant-Mode Inversion of Surface Waves: Inherently Addressing Inconspicuous Low Frequency Mode Jumps

Mrinal Bhaumik[*] and Brady R. Cox

Department of Civil and Environmental Engineering, Utah State University, Logan, Utah, USA

**Abstract**

Inversion of Rayleigh-wave dispersion data is particularly challenging at sites with strong impedance contrasts, where modal energy often transitions smoothly from the fundamental to higher modes at low frequencies. As a result, analysts can misinterpret the low-frequency portion of the dispersion data as a continuation of the fundamental mode, leading to overestimation of shear wave velocity ($V_s$) in deeper layers and/or misinterpretation of bedrock depth. Although effective-mode inversion formulations can theoretically account for these smooth modal transitions, their use requires precise knowledge of the source–receiver geometry and cannot be applied when target dispersion data are formed by combining dispersion data from multiple active shots with dispersion data extracted from passive-array recordings that have unknown source locations. This study introduces a predominant-mode inversion framework that inherently addresses low-frequency mode jumps by automatically identifying, at each frequency, the Rayleigh-wave mode with the maximum vertical surface amplitude. This allows inversions to proceed without the need for explicit mode indexing by the inversion analyst or unvalidated assumptions about fundamental modal dominance. The predominant-mode forward model is derived herein and implemented using the thin-layer method. Theoretical parametric analyses performed with this model have confirmed that modal osculation emerges when the subsurface velocity contrast exceeds approximately a factor of two. The predominant-mode forward model has been integrated into a particle-swarm-optimization global search algorithm and used to invert three synthetic models exhibiting low-frequency mode jump behavior. The inversion targets for synthetic models were generated through wavefield simulations mimicking active and passive surface wave testing. The inversions were performed with multiple layering parameterizations. Across all cases, the predominant-mode method accurately recovered major velocity contrasts and interface depths, whereas fundamental-mode inversions consistently overestimated $V_s$ and mislocated deeper layer boundaries. The method was further validated using real field data from active-source and passive-wavefield surface wave measurements at the I15 Downhole Array Site. Inverted $V_s$ profiles showed strong agreement with invasive downhole PS logs and with empirical transfer functions computed from recorded ground motions. Overall, the predominant-mode framework offers a robust and practical approach for surface-wave inversion at sites with strong impedance contrasts.

*Keywords:* Geophysics; Numerical modelling; Reliable subsurface models; Inversion; Modal osculation.

---

[*] Corresponding author
Email: mrinal.bhaumik@usu.edu

## 1.1. Introduction

Surface wave methods have emerged as one of the most widely applied non-invasive geophysical approaches for subsurface characterization due to their cost-effectiveness, applicability across diverse depth scales, and ability to provide information on the stiffness of soils and rocks without requiring invasive drilling. A key advantage of surface wave methods is their ability to infer in-situ shear wave velocity ($V_s$) profiles, which are fundamental for modeling the dynamic response of the ground to either seismic waves or anthropogenic loading (e.g., pile driving). In practice, surface wave testing is carried out using both active-source techniques, such as the Multichannel Analysis of Surface Waves (MASW) (Park et al., 1999; Foti, 2000), and passive-source techniques, such as Microtremor Array Measurements (MAM) (Okada, 2003; Foti et al., 2018). Active tests are effective for profiling shallow to intermediate depths using high to moderate frequencies (i.e., short to moderate wavelengths), while passive tests often allow for profiling to greater depths by using moderate to low frequencies (i.e., moderate to long wavelengths) (Garofalo et al., 2016; Foti et al., 2018). Both approaches require measuring the dispersive properties of surface waves across a broad range of frequencies/wavelengths. In regards to the propagation of surface waves, dispersion refers to a propagation velocity that is dependent on frequency/wavelength, or in other words a phase velocity. Surface waves are dispersive because higher frequencies/shorter wavelengths sample only shallow strata, while lower frequencies/longer wavelengths sample both shallow and deeper strata. While both Rayleigh- and Love-type surface waves are dispersive and can be used for surface wave testing, it is more common to use Rayleigh waves. By measuring Rayleigh phase velocity as a function of frequency, one can indirectly infer the variation of $V_s$ with depth. The key step in this process is the inversion of the experimental Rayleigh dispersion data, where the recorded phase velocity–frequency data is iteratively matched with theoretical Rayleigh dispersion curves computed for trial layered earth models. Conventionally, inversion is performed using the fundamental mode of Rayleigh waves, particularly at sites with monotonically increasing $V_s$ profiles, unless there is clear evidence of higher-mode excitation in the experimental dispersion data.

Higher modes of Rayleigh-type surface waves can be excited under two general conditions: (1) in the presence of a velocity reversal, such as low-velocity or high-velocity layers embedded within the profile, or (2) when there is a strong impedance contrast, for example, a soft sedimentary layer overlying dense sand or rock. In the case of low-velocity layers (LVLs), higher modes typically appear at higher frequencies with a distinct cutoff frequency, and they often continue exciting successively higher modes at higher frequencies, without reverting back to the fundamental mode (Mi et al., 2018). For embedded high-velocity layers (HVLs), the modal energy may transfer to a higher mode and then return to the fundamental mode, with the transition frequency depending on the depth of the HVL (O'Neill et al., 2003; Pan et al., 2019;

Bhaumik & Naskar, 2024b). In both cases, mode identification in the experimental dispersion data tends to be relatively straightforward, as the transitions are more abrupt and well-defined in the dispersion image, particularly when larger array apertures are used to record the data. In contrast, when higher modes arise due to a strong impedance contrast, they are typically excited at lower frequencies (Boaga et al., 2013). In these cases, the modal energy may gradually transition from the fundamental mode to a higher mode, making it difficult to detect in the experimental dispersion data. This gradual transition of modal dominance, often referred to as mode osculation or mode kissing, occurs when two modal branches approach each other closely in the dispersion curve (Boaga et al., 2013; Kausel et al., 2015; Gao et al., 2016). Unlike the abrupt mode separation observed at sites with velocity reversals, the transition of modal energy due to strong impedance contrast is often smooth and continuous due to lack of spatial resolution, making visual identification of the correct mode more challenging when examining the experimental dispersion data (Foti et al., 2018). The transition frequency is influenced by both the degree of the velocity contrast (velocity ratio) and the depth to the high-velocity interface (e.g., bedrock). This phenomenon introduces ambiguity in mode selection and may result in significant inversion errors, such as overestimation of deep-layer velocities or misinterpretation of bedrock depth.

The problem of mode misidentification can be addressed using concepts such as the effective mode, superimposed mode, or apparent mode (Lai, 1998; Maraschini et al., 2010; Foti et al., 2014, 2018; Lai et al., 2014). However, its computation requires explicit information about the source–receiver geometry, including source offset, receiver spacing, and array length (Lai et al., 2014). This makes it well suited for active-source MASW testing, where the source is placed at a controlled offset relative to the spread of receivers. However, in practice, active MASW and passive MAM data are often combined to extend the frequency range for deeper exploration, a strategy that is now quite common in many site characterization projects (Garofalo et al., 2016; Teague & Cox, 2016; Foti et al., 2018; Vantassel & Cox, 2022). In such cases, applying the effective-mode approach becomes difficult because the source locations are unknown in passive MAM surveys, and the inversion target is typically constructed by merging dispersion data from multiple MASW active shots with different source offsets (e.g., on both sides of the array) and several passive MAM arrays of different sizes (Vantassel & Cox, 2022). As such, there is a need for an effective mode inversion approach that is independent of rigorous source and receiver location information.

In this study, we propose a new, predominant-mode inversion strategy to mitigate errors arising from incorrect Rayleigh-wave mode assumptions. Instead of extracting only phase velocity, we also compute the complete mode shapes from the forward model and examine their amplitude distribution at the free surface, since that is where the measurements are recorded. By doing so, we can directly identify which mode contributes most strongly to the observable wavefield on the surface at each frequency in the dispersion data. The Rayleigh wave mode with the maximum vertical amplitude at the surface for each frequency, referred to as the predominant mode, is selected as the basis for inversion. The underlying idea is that an incorrectly assigned mode will inevitably lead to higher misfit, as its phase velocity does not match the recorded dispersion behavior and cannot adequately fit the experimental data. In contrast, the correctly extracted predominant mode from the forward model will align more closely with the recorded surface response and result in a lower misfit. Thus, by systematically choosing the mode that dominates the surface amplitude at each frequency, the inversion process avoids ambiguous mode numbering made by the inversion analyst and instead allows the data itself to inherently guide the selection of the relevant mode. Unlike the effective mode approach, which requires explicit knowledge of source offset, receiver spacing, and array geometry, the predominant mode does not depend on acquisition layout. As noted above, this independence makes it particularly well suited for situations where the inversion target is constructed from a combination of several MASW active shots with different offsets and/or when combining MASW dispersion data with that obtained from multiple passive MAM arrays, where source positions are unknown.

To investigate the applicability of a predominant mode inversion, the distribution of modal amplitude at the surface was first analyzed for a simple, two-layer system (one layer over half-space) to examine the conditions under which modal osculation arises. Different velocity contrasts at varying depths were considered in this parametric analysis. Following the parametric analysis, the predominant mode definition was integrated into a particle swarm optimization (PSO)–based global inversion framework to evaluate its performance. Three synthetic datasets and one real field dataset were employed in the evaluation. In all cases, the predominant mode successfully identified the modal transition frequencies and enabled development of reliable $V_s$ profiles. By relying on the predominant mode, the inversion framework avoids ambiguity in mode identification and reduces bias near transition frequencies. This approach is designed to enhance the robustness and reliability of $V_s$ profiling in stratigraphic settings characterized by significant impedance contrasts and complex layering.

## 2.1. Forward problem: Computation of modal amplitude and predominant mode

In this section, we briefly describe the methodology adopted to extract the predominant mode. The theoretical forward modeling was performed using the Thin Layer Method (TLM) (Kausel & Roësset, 1981). Among several available forward modeling approaches, the TLM offers a distinct advantage: it formulates the Rayleigh wave dispersion problem as an eigenvalue problem, allowing simultaneous computation of both the eigenvalues (phase velocities) and eigenvectors (mode shapes). This capability enables direct calculation of modal energy at the surface, which is essential for identifying the predominant mode. The detailed derivation and implementation of the TLM can be found in Kausel & Roësset, 1981; De Oliveira Barbosa & Kausel, 2012; and Bhaumik & Naskar, 2024b. It is important to note that this approach is general and can be adapted to any forward modeling scheme capable of producing mode shapes and phase velocities.

Considering a layered half-space, the governing equation of the dispersion relation for horizontally propagating Rayleigh waves can be expressed in quadratic eigenvalue form as (Kausel & Roësset, 1981):

$$\mathbf{K}\boldsymbol{\phi}_R = 0 \text{ , with } \mathbf{K}(\omega, k) = k^2 \mathbf{A} + ik\mathbf{B} + (\mathbf{C} - \omega^2 \mathbf{M}), \tag{1}$$

where $\boldsymbol{\phi}_R$ is the right eigenvector representing mode shape, $\omega$ is the angular frequency, $k$ is the wavenumber, and $i = \sqrt{-1}$. The global matrices $\mathbf{A}, \mathbf{B}, \mathbf{C}$ and $\mathbf{M}$ are assembled from the contributions of individual thin elements that discretize each soil layer along the depth axis. In this context, the x-components represent the horizontal in-plane displacement, while the z-components represent the vertical displacement, which together describe the coupled motion characteristic of Rayleigh waves. Arranging all horizontal components first, followed by the vertical ones, the global matrices are given as:

$$\mathbf{A} = \begin{bmatrix} \mathbf{A}_{xx} & 0 \\ 0 & \mathbf{A}_{zz} \end{bmatrix}; \mathbf{B} = \begin{bmatrix} 0 & \mathbf{B}_{xz} \\ -\mathbf{B}_{zx} & 0 \end{bmatrix}; \mathbf{C} = \begin{bmatrix} \mathbf{C}_{xx} & 0 \\ 0 & \mathbf{C}_{zz} \end{bmatrix}; \mathbf{M} = \begin{bmatrix} \mathbf{M}_{xx} & 0 \\ 0 & \mathbf{M}_{zz} \end{bmatrix} \tag{2}$$

At the element level, the corresponding sub-matrices are defined as,

$$\mathbf{A}_{xx}^{(e)} = (\lambda + 2\mu) \int \mathbf{N}^T \mathbf{N} \, dz, \mathbf{A}_{zz}^{(e)} = \mu \int \mathbf{N}^T \mathbf{N} \, dz, \mathbf{B}_{xz}^{(e)} = \lambda \int \mathbf{N}^T \mathbf{N}' \, dz - \mu \int \mathbf{N}'^T \mathbf{N} \, dz, \mathbf{B}_{zx}^{(e)}$$
$$= \mathbf{B}_{xz}^{(e)T}, \mathbf{C}_{xx}^{(e)} = \mu \int \mathbf{N}'^T \mathbf{N}' dz, \mathbf{C}_{zz}^{(e)} = (\lambda + 2\mu) \int \mathbf{N}'^T \mathbf{N}' dz, \mathbf{M}_{xx}^{(e)} = \mathbf{M}_{zz}^{(e)} \tag{3}$$
$$= \rho \int \mathbf{N}^T \mathbf{N} \, dz$$

Here, $\mathbf{N}(z)$ denotes the interpolation (shape) functions and $\mathbf{N}'(z)$ their derivatives, $\rho$ is density, and $\lambda, \mu$ are the Lame's parameters. The quadratic eigenvalue problem in Equation (1) can be reformulated as a generalized eigenvalue problem of the same size (Kausel, 2005):

$$\left( k^2 \underbrace{\begin{bmatrix} \mathbf{A}_{xx} & 0 \\ \mathbf{B}_{xz}^T & \mathbf{A}_{zz} \end{bmatrix}}_{\widetilde{\mathbf{A}}} + \underbrace{\begin{bmatrix} \mathbf{C}_{xx} - \omega^2 \mathbf{M}_{xx} & \mathbf{B}_{xz} \\ 0 & \mathbf{C}_{zz} - \omega^2 \mathbf{M}_{zz} \end{bmatrix}}_{\widetilde{\mathbf{C}}} \right) \begin{Bmatrix} \boldsymbol{\phi}_x \\ ik\boldsymbol{\phi}_z \end{Bmatrix} = 0 \tag{4}$$

where $\boldsymbol{\phi}_x, \boldsymbol{\phi}_z$ are horizontal and vertical displacement components of the right eigenvector, $\widetilde{\mathbf{A}} = \begin{bmatrix} \mathbf{A}_{xx} & 0 \\ \mathbf{B}_{xz}^T & \mathbf{A}_{zz} \end{bmatrix}$, and $\widetilde{\mathbf{C}} = \begin{bmatrix} \mathbf{C}_{xx} - \omega^2 \mathbf{M}_{xx} & \mathbf{B}_{xz} \\ 0 & \mathbf{C}_{zz} - \omega^2 \mathbf{M}_{zz} \end{bmatrix}$. The above equation can be expressed in compact matrix notation as:

$$\Lambda^2 \widetilde{A} R + \widetilde{C} R = 0 \quad (5)$$

where, $\Lambda = \text{diag}\{k_1^2, k_2^2, \dots\}$ is the matrix containing the eigenvalues, and $R$ is the corresponding matrix of right eigenvectors. Since the matrices are nonsymmetric, distinct left and right eigenvectors must be considered. Following the orthogonality relations described in De Oliveira Barbosa & Kausel (2012), the eigenvectors are normalized such that they satisfy:

$$L^T \widetilde{A} R = \Lambda^{1/2} \text{ and } L^T \widetilde{C} R = -\Lambda^{3/2} \quad (6)$$

where $L$ is the left eigenvector matrix.

The solution of the eigenvalue problem in Equation (4) produces both real and complex roots. Because the focus here is on undamped, purely propagating Rayleigh-type surface waves, only the real roots are considered, and the corresponding normalised eigenvectors $\widehat{\boldsymbol{\phi}}_R = [\widehat{\boldsymbol{\phi}}_x \quad \widehat{\boldsymbol{\phi}}_z]^T$. At the free surface ($z = 0$), the normalised vertical component of mode $m$ is written as: $\widehat{\boldsymbol{\phi}}_{z,m}(0; k_m, \omega)$, which represents the theoretical surface amplitude associated with that mode at frequency $\omega$. By evaluating these amplitudes for all modes, the predominant mode is defined as the one with the largest vertical amplitude at the surface (Naskar & Kumar, 2017; Bhaumik & Naskar, 2024b),

$$k_{predominant}(\omega) = \arg\max_{k_m} |\widehat{\boldsymbol{\phi}}_{z,m}(0; k_m, \omega)| \quad (7)$$

The detailed derivation above establishes the key steps to identify the predominant mode. For clarity, this workflow can also be represented in a single compact expression that traces the sequence from the governing matrices, through the eigenvalue problem (EVP) and normalization, to the evaluation of surface amplitudes and the final mode selection:

$$(\widetilde{A}, \widetilde{C}) \xrightarrow[\text{EVP}]{\Lambda^2 \widetilde{A} R + \widetilde{C} R = 0} (k_m^2, \boldsymbol{\phi}_R) \xrightarrow[\text{Normalization}]{L^T \widetilde{A} R = \Lambda^{1/2}, L^T \widetilde{C} R = -\Lambda^{3/2}} \widehat{\boldsymbol{\phi}}_R \xrightarrow[\text{Surface amplitude}]{z=0} \widehat{\boldsymbol{\phi}}_{z,m}(0; k_m, \omega)$$
$$\xrightarrow{\arg\max} k_{predominant}(\omega) \quad (8)$$

The surface amplitude, $\widehat{\boldsymbol{\phi}}_{z,m}(0; k_m, \omega)$, will be used in the subsequent section for further investigations into the physical conditions under which modal transitions occur and how these are influenced by subsurface structure.

## 3.1. Modal Osculation and Velocity Contrast

### 3.1.1. Influence of Velocity Contrast and Depth

In this section, we examined how velocity contrast and its corresponding depth influence the distribution of modal energy at the surface. For this purpose, a simple two-layer model was considered, consisting of a top layer with $V_s$ = 100 m/s, underlain by an elastic half-space with varying shear-wave velocity contrasts (VC). Among several trial cases, four representative cases with VC = 2, 3, 5, and 8 are reported here. Each VC case was analyzed for three different top-layer thicknesses (10 m, 20 m, and 30 m) to evaluate the effect of depth. Figure 1 presents the corresponding outcomes of this parametric study for the vertical component of Rayleigh waves. The theoretical dispersion curves are superimposed on the modal energy maps, which were computed from the amplitude of the normalized eigenvectors (mode shapes) at the surface, ($\hat{\phi}_{z,m}(0; k_m, \omega)$), obtained from Equation 7. The results clearly indicate that, irrespective of depth, when the VC exceeds approximately 2, the dominant modal energy shifts from the fundamental Rayleigh mode (R0) to the first-higher Rayleigh mode (R1). This energy transfer is a key signature of strong impedance contrast and is consistent with the phenomenon often described in the literature as mode kissing or modal osculation (Boaga et al., 2013; Kausel et al., 2015; Gao et al., 2016; Yust et al., 2018). The frequency at which the transition between the R0 and R1 occurs is referred to as the osculation point (OP). The results show that increasing depth to the velocity contrast does not fundamentally alter the shape of the dispersion curve or the corresponding modal energy distribution; instead, it shifts the entire pattern toward lower frequencies, consistent with the longer wavelengths required to explore deeper interfaces. Interestingly, for a VC of 8, the fundamental mode carries most of the modal energy across the observed frequency range (Figure 1d, h, and l). When the VC becomes large, on the order of 12 or greater (not shown here), the fundamental mode can dominate the entire frequency spectrum, effectively suppressing higher-mode excitation. However, this threshold value is not universal; it depends on other factors, such as the overall velocity profile, the number of layers, and their thicknesses. Importantly, the required velocity contrast does not have to occur between two immediately adjacent layers. It can develop gradually through several thin layers within a relatively small depth interval, ultimately producing the same modal energy transfer behavior. This has important implications for real geological profiles, where sharp, single-layer interfaces are less common than gradual transitions. The return of modal energy from R1 back to the R0 mode (see the lower-frequency portion for VC values of 3, 5, and 8) is governed entirely by the velocity structure. In synthetic profiles, a sharp, single, high-velocity layer abruptly truncated at a certain depth and directly underlain by a half-space can produce a distinct reversion. However, such idealized configurations are geologically uncommon. In most real-world stratigraphy, additional layers or gradual velocity increases occur below the strong contrast, which can significantly alter or even suppress the return of R0. For this reason, the present discussion focuses on the initial jump of modal energy from R0 to R1, as this behaviour is consistently observed.

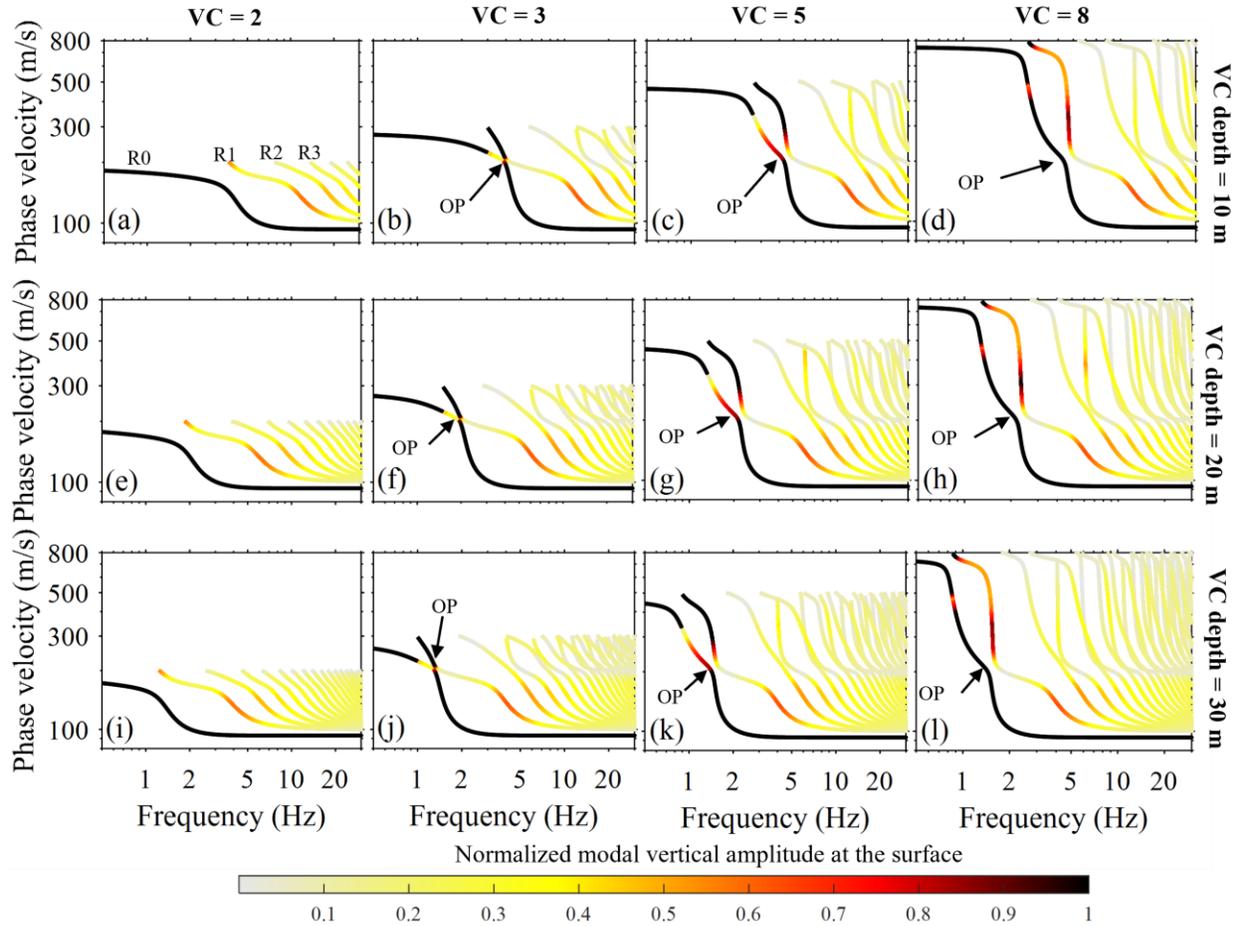

**Figure 1.** Theoretical Rayleigh wave dispersion curves colored according to normalized modal vertical surface amplitudes, illustrating the effect of velocity contrast (VC = 2, 3, 5, and 8) and contrast depth (10, 20, and 30 m) on modal behaviour.

Although only the vertical component of Rayleigh waves is shown here, the radial component exhibits similar, and sometimes more abrupt modal energy transitions. In contrast, Love waves do not exhibit any significant modal energy transfer to higher modes due to velocity contrast; the predominant modal energy consistently remains in the fundamental mode, even though a larger number of higher-mode branches are excited compared to Rayleigh waves. Since the vertical component of Rayleigh waves is the most commonly used in current practice, the present discussion focuses exclusively on this component. Importantly, in field data, the transition of modal energy between R0 and R1 is often smooth and gradual, primarily due to insufficient spatial resolution caused by recording the wavefield with relatively short arrays with relatively large receiver spacing, making the distinct transition between fundamental and higher modes less obvious in experimental dispersion images. The absence of a sharp visual signature makes mode identification difficult, and this uncertainty can lead to velocity overestimation during inversion when R1 is accidentally fit as if it were R0 after the osculation point.

### 3.1.2. Consequence of Mode Misidentification

The practical implication of such gradual modal transitions is that the experimental dispersion data may be incorrectly attributed to the fundamental branch by the inversion analyst, leading to velocity overestimation or the derivation of biased subsurface profiles. As an illustrative example, we performed a wavefield simulation for the two-layer model with VC of 3 at a depth of 20 m. Recall, the mode-amplitude weighted dispersion curves for this model are shown in Figure 1f. The simulation was performed using the (2-4) staggered grid finite difference method (Virieux, 1986; Bhaumik & Naskar, 2023), a numerical scheme that discretizes stress and particle velocity components on an offset grid to achieve second-order accuracy in time and fourth order in space. The simulated wavefield was recorded using an array of 150 receivers spaced at 2-m intervals. For visualization purposes, only every 10th receiver is shown in Figure 2a, with the first receiver located at a source offset of 10 m. The wavefield was processed using the standard f-k method (Yilmaz, 1987) with spectral whitening (Naskar et al., 2025) and the resulting dispersion image (Figure 2b) reveals how the modal energy smoothly transitions from the fundamental mode to the first higher mode at the osculation point around 2 Hz, without clear separation between modes. This smooth modal energy transition introduces a significant challenge for analysts, as it obscures the boundary between modes and makes visual identification by the inversion analyst unreliable. The error bars indicate the peak-power dispersion data points obtained from the dispersion image with +/- one standard deviation uncertainty bounds assigned based on common experimental dispersion data coefficients of variation equal to 5% (Cox et al., 2014; Garofalo et al., 2016). To illustrate the practical impact of not recognizing this dispersion data consists of multiple modes, Figures 2c and 2d present the inversion results obtained by interpreting the peak dispersion data as purely fundamental mode. The inversion was performed using the *Dinver* toolbox in *Geopsy* (Wathelet, 2008). Even though the true model was known to consist of two layers, several layering configurations were used as a means to investigate inversion sensitivity to the choice of layering parameterization. Trial models consisting of two, three, and four layers were used for the inversions. Figure 2c shows the target dispersion data from Figure 2b and the overall 100 best-fit theoretical dispersion curves obtained from fitting the experimental data under the R0 assumption. The inversions produced the two-, three-, and four-layer $V_s$ profiles shown in Figure 2d, all of which substantially overestimate the velocity and depth of the half-space in comparison to the true model. This overestimation is a direct consequence of incorrectly assuming the fundamental mode dominates the entire frequency range of the dispersion data, highlighting how smooth modal energy transitions caused by strong impedance contrasts and can lead to systematic errors in estimating deep subsurface velocities or bedrock depth.

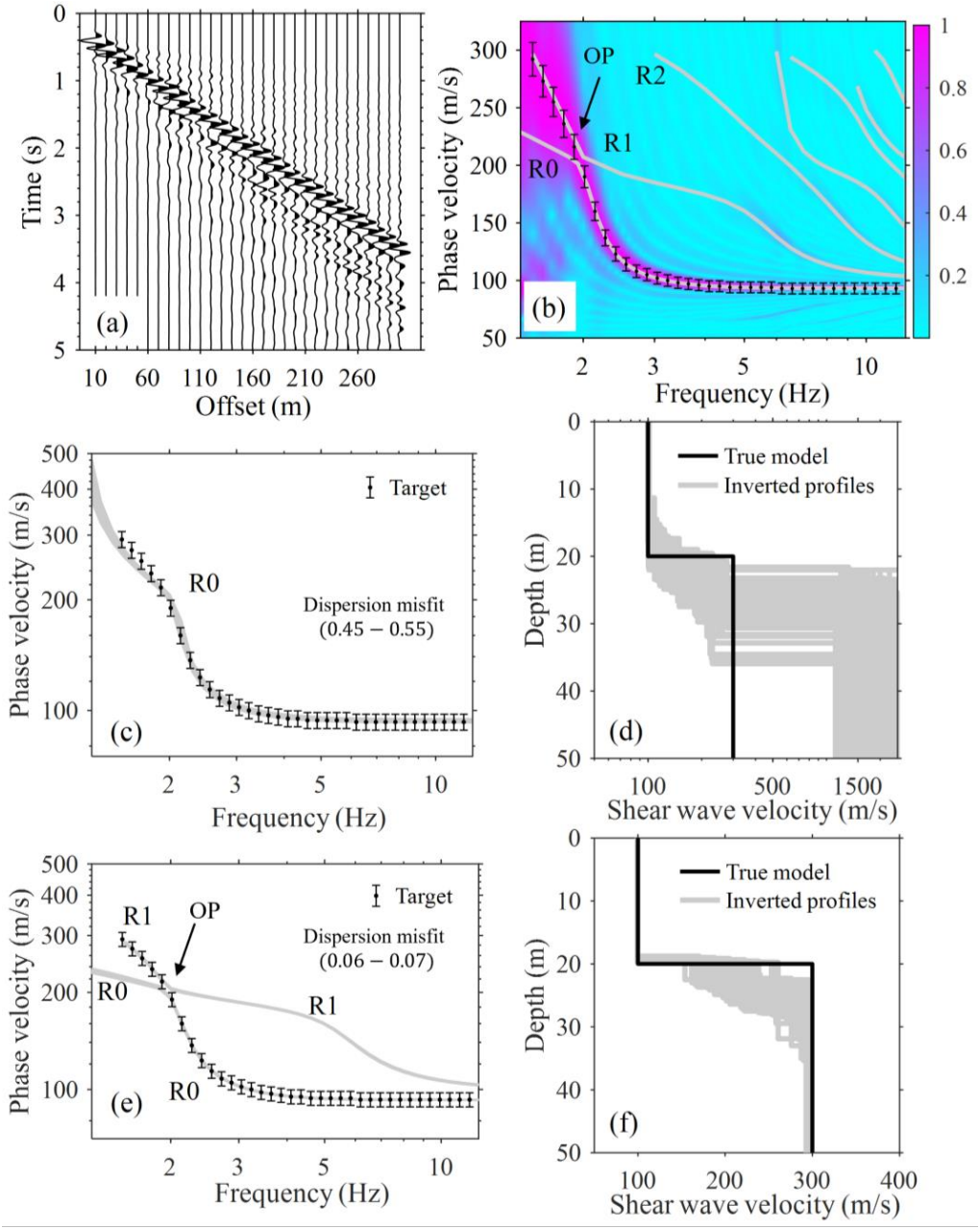

**Figure 2.** Impact of mode misidentification on surface-wave inversion: (a) simulated active-shot waveforms, (b) corresponding dispersion image highlighting the smooth transition of modal energy from R0 to R1, with the observed (target) picks indicated, (c) the 100 best-fitting theoretical dispersion curves obtained when the target picks are incorrectly assigned to R0, and (d) the corresponding inverted $V_s$ profiles, which systematically overestimate the half-space depth and velocity compared to the true model, (e) the 100 best-fitting theoretical dispersion curves obtained when the target picks are correctly associated with the appropriate R0 and R1 modal index, and (f) the resulting inverted $V_s$ profiles, which closely reproduce the true model.

On the other hand, if a multimodal inversion is performed by considering the simulated dispersion data above 2 Hz as R0 and that below 2 Hz as R1, the inversion results are shown in Figure 2e and 2f. For this case, the inversions were also performed in *Dinver* (Wathelet, 2008) after breaking the simulated dispersion data into two parts at the osculation point and specifying the higher frequency target as R0 and the lower frequency target as R1. It can be observed that with correct and manual mode identification by the analyst, the inversion can retrieve $V_s$ profiles that are much closer to the true profile, even when uncertainty in layering parameterization is considered. However, the analyst must manually identify the exact point of modal energy transition and use an inversion algorithm cable of multimodal inversion.

This numerical example demonstrates that smooth modal transitions at low frequenceis make visual interpretation of the dispersion data unreliable and can lead to systematic overestimation of depth and shear-wave velocity when the dispersion data are incorrectly attributed to R0. These challenges highlight the need for an inversion strategy that does not rely on subjective mode numbering based on visual inspection by the analyst prior to inversion. To address this limitation, the following section introduces the predominant-mode inversion framework, in which the mode carrying the theoretical maximum surface energy at each frequency is automatically identified from forward modeling and used as the basis for inversion.

## 4.1. Predominant-Mode Inversion Framework

The predominant-mode calculated using Equation 7 was embedded within an inversion workflow. The framework is organized as a sequence of steps, beginning with the selection of representative synthetic models, followed by the development of target dispersion data, model parameterization, and the inversion procedure. Results from synthetic case studies are then presented to evaluate the performance of the approach before applying it to field data.

### 4.1.1. Synthetic Model Selection

To evaluate the performance of the predominant-mode inversion, three synthetic ground models were selected from the 12 synthetic models publish by Vantassel & Cox (2021) when developing their inversion procedure called *SWinvert*. Specifically, ground Models g, h, and k were chosen for this study because they were categorized by Vantassel and Cox (2021) as having significant velocity contrasts with "L-shaped" dispersion curves (i.e., rapidly increasing phase velocity at low frequencies/long wavelengths) that resulted in inverted $V_s$ profiles that did not accurately match the true synthetic $V_s$ profiles (i.e., had high $V_s$ misfit values). The $V_s$ profiles for Models g, h, and k are shown in Figure 3a-c along with their theoretical dispersion curves in Figure 3d–f, which are colored according to the normalized modal vertical amplitude at the surface. As seen in all three models, the presence of sharp impedance contrasts results in the excitation of higher modes at low frequencies. The OP, where the modal energy transitions from the R0 to R1 mode, is indicated in the figure. To generate target dispersion data for inversion, full wavefield simulations were performed to replicate both active-source MASW and passive-source MAM surface-wave testing typically conducted in the field. Phase velocities extracted from multiple active shots and various passive array apertures are shown in the lower panels of Figure 3 as experimental dispersion data, with the associated statistical variability also indicated (Figure 3d–f). Details of the wavefield simulation process and the methodology used to extract these experimental dispersion data are provided in the following section.

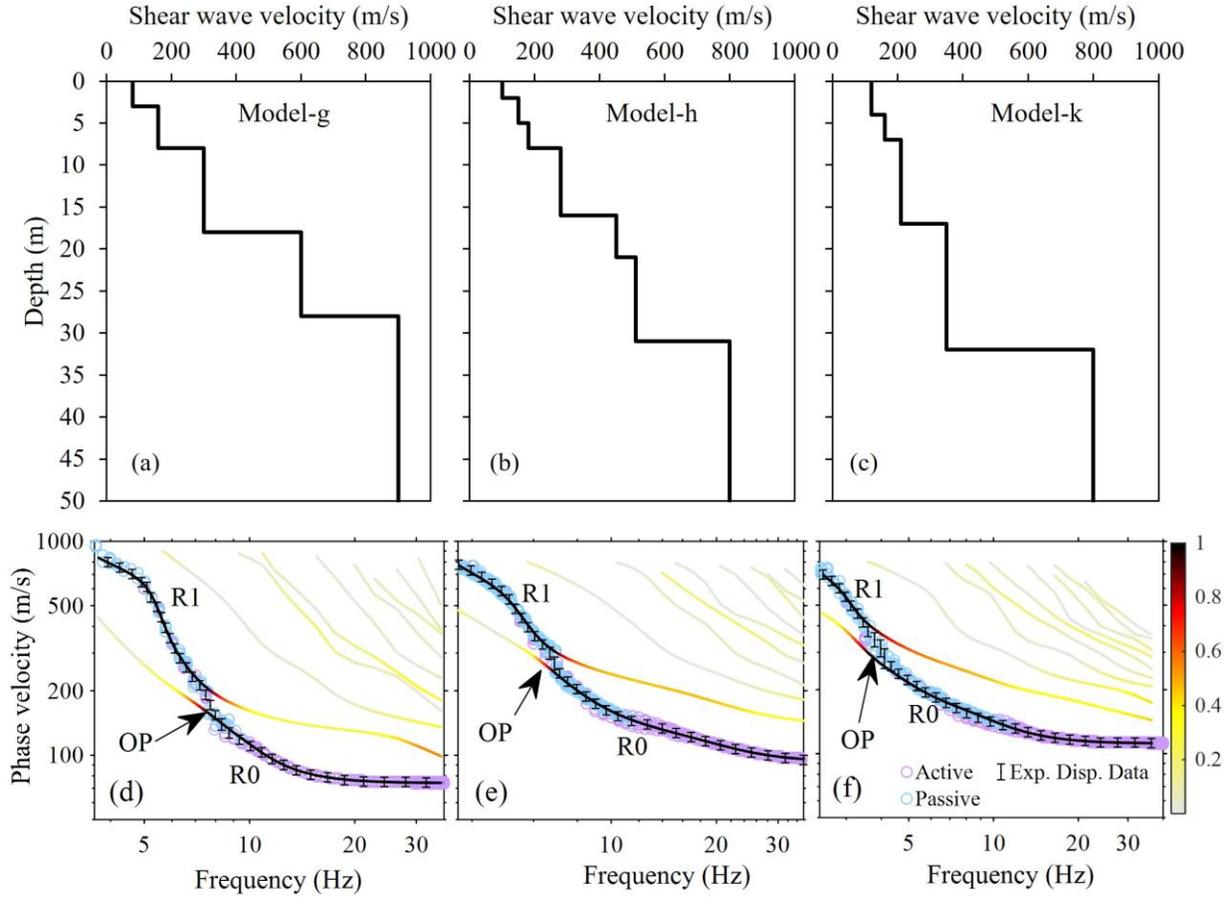

**Figure 3.** Shear-wave velocity profiles for the three synthetic ground models adopted from Vantassel & Cox (2021) and used as inversion targets in this study: (a) Model g, (b) Model h, and (c) Model k. Panels (d)–(f) show the corresponding theoretical dispersion curves for each model with normalized modal vertical surface amplitudes together with the simulated active and passive dispersion measurements. The merged experimental dispersion data , formed by combining the active and passive measurements, are plotted as error bars to represent the full dispersion data set used in the inversion process.

### 4.1.2. Development of Target Dispersion Data

To develop the target experimental dispersion data for Models g, h, and k, both active MASW and passive MAM surface wave tests were simulated using the wavefield modelling method described by Bhaumik &

Naskar (2024a). The simulation method operates in the frequency domain, recording the response of specific frequencies at selected surface locations, and the time-domain response is subsequently obtained via inverse Fourier transform. For the active test, 24 receivers were placed in a linear array at 1 m intervals, with source offsets of 5, 10, 15 and 20 m, as illustrated in Figure 4a. An example active shot record with 5 m source offset is shown in Figure 4b. For the passive test, 10 receivers were deployed in circular arrays, with nine positioned along the circumference and one at the centre. Arrays with diameters of 50 m and 100 m were used. The passive wavefield simulations were performed by distributing several virtual sources across the study area, each with a randomly assigned phase and amplitude to mimic the incoherent nature of real ambient vibrations. The response at each receiver was then computed separately for each source, similar to an active test. Afterward, the recorded responses from all sources were superimposed, producing a synthetic wavefield that approximates the cumulative effect of ambient excitations. In the present study, 100 sources were randomly placed between 2000 m and 4000 m, each with a random phase delay to mimic natural, far-field ambient vibrations (Figure 4c). An example of the passive waveforms is shown in Figure 4d. Wavefield transformation to obtain dispersion data was then performed using frequency-domain beamforming (Zywicki & Rix, 2005) for the active simulation and high-resolution f-k for the passive simulation (Wathelet, 2008). Finally, the target dispersion data were extracted using the *SWprocess* tool, which supports calculating dispersion data uncertainty and sets a minimum coefficient of variation on phase velocity of 5% (Vantassel & Cox, 2022). The resulting dispersion data provide realistic representations of surface wave propagation, capturing both fundamental and higher-mode contributions (Figures 3d-f). These dispersion data serve as target data for the subsequent predominant mode inversion.

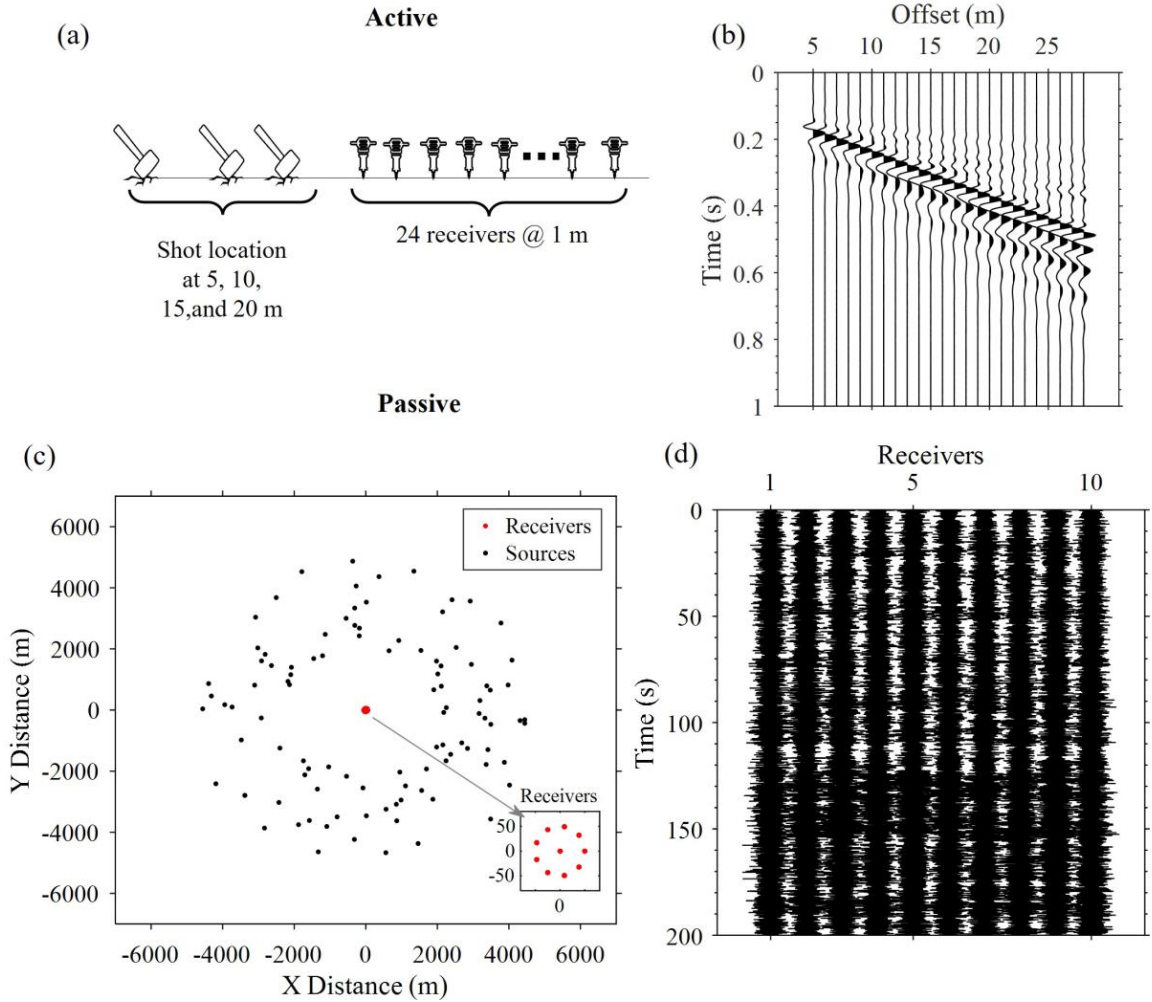

**Figure 4.** Generation of synthetic dispersion data for inversion: (a) active-source acquisition layout showing four shot positions at 5, 10, 15, and 20 m offset from a 24-channel linear array with 1 m spacing; (b) example of a simulated active shot gather for the 5 m source offset; (c) passive-source simulation using a spatially distributed set of noise sources and a circular array with 10 receivers (enlarged inset shows receiver geometry); and (d) example of simulated passive records.

### 4.1.3. Model Parameterization

With the inversion targets established, the next step is to define a suitable parameterization of the subsurface model. An appropriate parameterization of the subsurface, specifically, the number of layers, their thicknesses, and their ranges of $V_s$ and P-wave velocities ($V_p$), is critical for reliable inversion. When a priori information such as borehole logs, geophysical boundaries, or water table depth is available, it should be used to constrain model parameters and reduce epistemic uncertainty. However, in the absence of such information, careful selection of the number of layers and their thicknesses becomes essential. In surface wave inversion, the number of layers, their thicknesses, and associated $V_s$ are typically the most influential parameters controlling the shape and sensitivity of the theoretical dispersion curves. A common but simplistic approach for inversion parameterization is to assume a fixed number of equally thick layers up to the estimated resolution depth (typically assumed as ½ the maximum experimental wavelength) and to allow only $V_s$ to vary. While straightforward, this can result in over-parameterization, especially if the number of layers exceeds the resolution capability of the data, leading to non-uniqueness and poor convergence of the optimization algorithm. Additionally, excessive layering tends to over-smooth the inverted profile, obscuring sharp velocity contrasts such as the interface between soft sediments and bedrock. However, adopting a layering scheme with increasing thickness at greater depths is consistent with the physics of surface waves, which lose their ability to resolve thin layers with increasing depth due to vertical averaging of the wavefield. Cox & Teague (2016) proposed a layering ratio (LR) approach, which encourages, but does not force, increasing layer thickness with depth, while simultaneously providing a systematic method for selecting and managing multiple layering schemes in an attempt to account for subsurface layering uncertainty during inversion (Vantassel & Cox, 2021). The inversion parameterizations in this study were developed using the LR approach. The thickness of the first layer was constrained between one-half and one-third of the minimum observed wavelength, ensuring sufficient resolution of the shallowest structure. The potential thickness of each subsequent layer was increased with depth based on the selected layering ratio. Rather than selecting a single layering configuration, the LR approach explores a range of LR values, typically from 1.5 to 5, resulting in several candidate subsurface layering discretizations. Smaller LR values result in model parameterizations with more layers and smaller impedance contrasts, while larger LR values result in model parameterizations with fewer layers and larger impedance contrasts. This strategy is particularly advantageous when no a priori stratigraphic information is available, as it encourages exploration of multiple subsurface model structures. For each candidate model, the allowable $V_s$ range for each layer was selected based on the observed phase-velocity dispersion data. The minimum $V_s$ was set to one-half of the minimum observed phase velocity, and the maximum $V_s$ was initially limited to 1000 m/s; however, this upper bound was increased when ensembles of trial profiles began to reach the limit. The mass density was fixed at 2000 kg/m³ for all layers, and the Poisson's ratio was permitted to vary between 0.2 and 0.35. The details of the specific parameterizations adopted for the inversion procedure are presented in the following section.

### 4.1.4. Inversion Procedure

Global search inversion was performed using Particle Swarm Optimization (PSO) implemented through MATLAB's built-in Global Optimization Toolbox. Although the native *particleswarm* function was used, several key optimization parameters were carefully selected to balance exploration and exploitation of the model space. The optimization involved multiple unknowns, including layer thicknesses, $V_s$, and Poisson's ratios. Appropriate upper and lower bounds for each parameter were defined as described in the previous section. Given the high dimensionality of the problem, a swarm size of 150 particles was selected. This means that in each iteration, 150 candidate models are evaluated (i.e., 150 forward simulations per iteration), and the algorithm tracks both the global best and individual particle best models based on the misfit values. To encourage broad exploration of the solution space and ensure convergence, each PSO trial was divided into four consecutive stages: the first stage used 100 iterations with a *MinNeighborsFraction* of 0.5, an inertia weight range of 0.6–1.0, and social and cognitive adjustment weights set to 1.5. These values encourage diverse particle motion and prevent the entire swarm from prematurely clustering around local solutions, which is critical during early exploration. The final profile with minimum misfit serves as the initial or starting model for the subsequent stages. The second and third stages used 50 iterations each with *MinNeighborsFraction* set to 1.0 and a reduced inertia range of 0.5–0.8 to gradually shift toward exploitation. The final stage involved 100 iterations, continuing with an exploitation-focused setup. Altogether, each trial ran for 300 iterations, totalling 45,000 forward simulations per trial; across 10 trials, approximately 450,000 forward models were computed. Since MATLAB's PSO does not inherently support non-linear constraints, such as limiting velocity reversals or enforcing a maximum half-space velocity, a model rejection criterion was included in the objective function by penalizing models that violated acceptability, particularly those with non-monotonic velocity profiles. Specifically, any model exhibiting velocity reversals beyond a predefined threshold constraint was assigned with a large misfit penalty, effectively removing it from the search. While this initially led to frequent model rejections, the swarm quickly adapted and concentrated its search within the feasible region, reducing rejections and enhancing convergence in later iterations.

The inversion objective function was defined as a weighted root-mean-square dispersion misfit ($m_d$) value between the observed dispersion data and modelled dispersion curves (Wathelet, 2005):

$$m_d = \sqrt{\sum_{i=1}^{N_f} \frac{(\boldsymbol{V}_{\text{obs},i} - \boldsymbol{V}_{\text{predominant},i})^2}{N_f \sigma_i^2}} \qquad (9)$$

where $\boldsymbol{V}_{\text{obs},i}$ is the observed phase velocity at frequency $i$, $\boldsymbol{V}_{\text{predominant},i}$ is the modeled predominant-mode velocity at the same frequency, $N_f$ is the total number of frequency points, and $\sigma_i$ is the standard deviation associated with the observed data at frequency $i$.

## 5.1. Results of Synthetic Examples

Following the approach of Vantassel and Cox (2021), this study considers LRs of 1.5, 2, and 3, which are among the most commonly used in practice. Although their work explored a wider range of LRs (1.2, 1.5, 2, 3, 5, and 25), some were excluded for practical reasons. Specifically, an LR of 1.2 produced a large number of layers, making the model cumbersome, LR = 5 was rejected in their study as it resulted in very high misfit values, and LR = 25 is excessively coarse. While Vantassel and Cox (2021) also discussed the layering by number (LN) inversion parameterization approach, only the LR method is applied here for simplicity.

It is important to highlight that the inversion framework applied by Vantassel and Cox (2021) differs from the present study. In their work, the theoretical fundamental-mode dispersion curves (i.e., the R0 curves in Figure 3) were used to create target dispersion data with typical 5% COV error bars, and the inversion was carried out using only the fundamental mode across multiple parameterizations and independent trials. Their results, therefore, represent the best possible fundamental-mode inversion benchmark, considering only dispersion uncertainty and inversion non-uniqueness. In contrast, the present study uses dispersion data extracted from wavefield simulations as the targets, mimicking the actual dispersion data one would expect to extract from active and passive tests conducted in the field. Therefore, to demonstrate the consequences of mode misidentification (i.e., incorrectly assuming that the observed dispersion data corresponds to the fundamental mode), the results from a conventional fundamental-mode inversion of the simulated experimental dispersion data (i.e., the error bars in Figure 3) are included for comparison. Fundamental-mode inversions were performed using the *Dinver* toolbox within *Geopsy*. Following the procedure by Vantassel and Cox (2021), each selected LR model parameterization was run for 10 independent trials. From these, both the single best-fitting profile and the 10 best profiles from each trial were retained for plotting and comparison. To quantitatively assess inversion accuracy, Vantassel and Cox (2021) computed shear wave velocity misfit ($m_{V_s}$) values between the true synthetic $V_s$ profiles and the inverted $V_s$ profiles up to a depth of 50 m, with a depth discretization of 0.5 m. Therefore, in this paper, $m_{V_s}$ values were calculated in the same manner for both predominant and fundamental mode-based inversion results and compared with the published $m_{V_s}$ values from Vantassel and Cox (2021) as a reference to provide a benchmark for evaluating the performance of the new predominant-mode inversion approach. The $m_{V_s}$ was calculated using mean absolute relative error (Vantassel & Cox, 2021):

$$m_{V_s} = \frac{1}{N} \sum_{j=1}^{N_d} \frac{\left|V_{s_j}, \text{inverted} - V_{s_j}, \text{true}\right|}{V_{s_j}, \text{true}} \tag{10}$$

where $V_{s_j}$, inverted and $V_{s_j}$, true are the inverted and true shear-wave velocities at depth point $j$, respectively, and $N_d$ is the number of discretization points.

An additional relative dispersion misfit ($m_{dr}$) value is required for the predominant-mode inversion framework. While the overall goodness of fit between theoretical predominant-mode dispersion curves and the simulated/observed dispersion data are calculated using Equation 9, a frequency-by-frequency relative dispersion misfit value is also of interest for the reasons discussed below. The frequency-by-frequency relative dispersion misfit between the theoretical predominant-mode phase velocity of the inverted profiles and the observed phase velocity is calculated via Equation 11:

$$m_{dr} \text{ in \%} = 100 \times \frac{|V_{obs,i} - V_{predominant,i}|}{V_{obs,i}} \tag{11}$$

The $m_{dr}$ helps in identifying individual frequency points where the theoretical predominant-mode is unable to adequately fit the target dispersion data. These points will have relatively high $m_{dr}$ values, allowing them to be examined or treated separately during subsequent/revised inversions. One may also use Equation 9 on a per-frequency basis without summation, however, normalizing the difference by $V_{obs}$ and expressing it in percentage, as in Equation 11, ensures that the misfit at each frequency is placed on a comparable scale. The appropriate use of $m_{dr}$ is illustrated in the synthetic model inversion case studies discussed below.

The following subsections present detailed results for the three selected synthetic Models (g, h, and k) shown in Figure 3, highlighting their specific challenges and the performance of the predominant-mode inversion.

### 5.1.1. Ground Model g

Model g is a 4-layer model with $V_s$ boundaries at 3, 8, 18, and 28 m depth, underlain by a half-space. Although the maximum velocity contrast at any single boundary is 2, the cumulative effect of successive contrasts leads to the excitation of higher modes. Figure 5 presents the results of the predominant-mode inversion alongside those obtained from a conventional fundamental-mode inversion of the simulated experimental dispersion data. Figure 5a presents the target experimental dispersion data, together with the theoretical R0 and R1 curves corresponding to the ten best-fitting models from each trial (100 curves per LR) for all three layering-ratio parameterizations. The colored curves represent the predominant-mode inversion results, whereas the gray curves represent the fundamental-mode inversion results. Similarly, the mode indices (i.e., R0 and R1) are also marked as gray for fundamental mode inversion and in black for predominant-mode inversion. All sets of predominant-mode inversion theoretical curves correctly track the transition from the R0 to the R1 mode near the OP at approximately 8 Hz. This demonstrates that the predominant-mode inversion inherently fits the energy-dominant branch of the dispersion curves. The range of $m_d$ values (obtained by Equation 9) from all 100 profiles per LR (reported at the bottom of Figure 5 inside brackets) confirm consistent performance of the predominant mode inversion across parameterizations; LR = 1.5, 2, and 3 yield comparable $m_d$ ranges, with LR = 3 showing slightly broader variability. In contrast, the fundamental-mode inversion (gray curves) follows the target dispersion data reasonably well in low and higher frequencies but begins to deviate noticeably between 8–10 Hz. This mismatch results in substantially higher $m_d$ values (1.21-1.54, depending on LR), which clearly shows that the dispersion data cannot be fit as well with a fundamental mode.

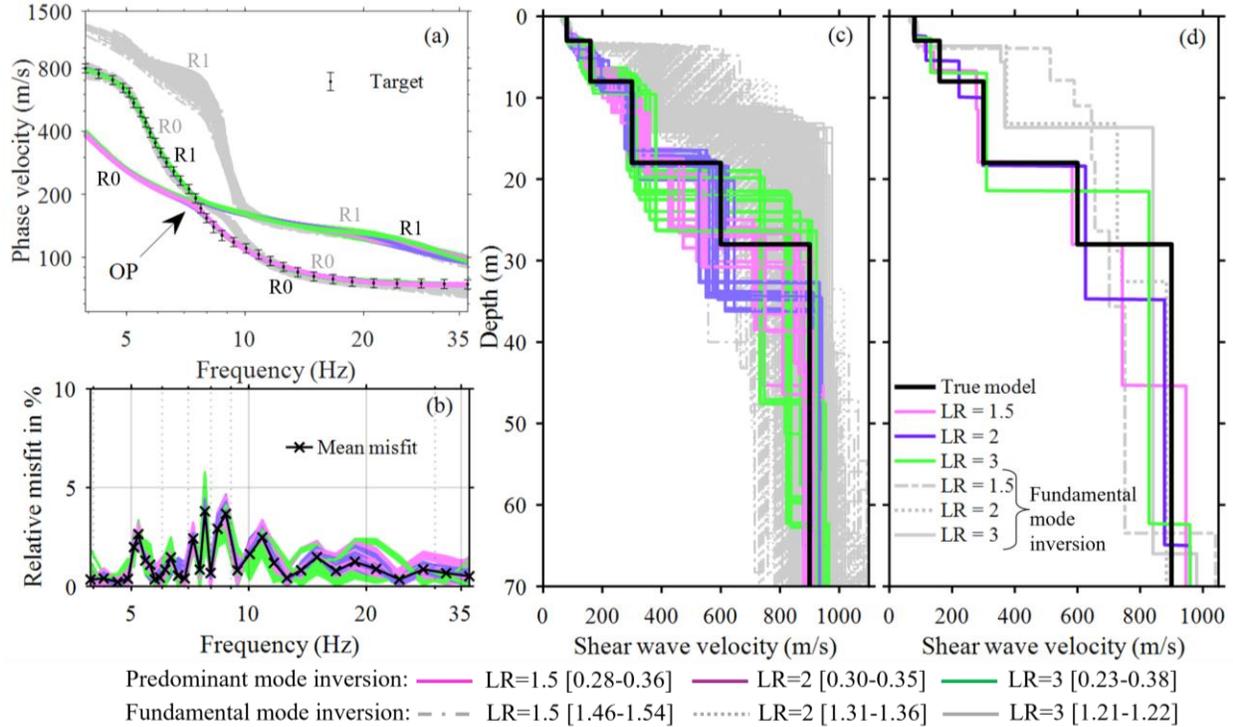

**Figure 5.** Inversion results for Model g: (a) dispersion data (black points with error bars) and inversion-derived theoretical dispersion curves for the ten best-fit solutions from each trial (100 models per LR); colored curves correspond to predominant-mode inversion; gray curves correspond to fundamental-mode inversion; (b) frequency-wise relative misfit between the theoretical dispersion curves and observed dispersion data, with the black line showing the mean misfit trend across all trials ; (c) corresponding inversion-derived $V_s$ profiles obtained from the same ten best-fit solutions for each trial (100 models per LR); and (d) best-fitting profiles from each predominant-mode LR inversion compared with the best fundamental-mode inversion results. Misfit values shown in brackets with each LR denote the range in dispersion misfits for the corresponding set of 100 models.

To more closely examine the fit of the predominant-mode inversion across the usable frequency range, Figure 5b plots the frequency-wise $m_{dr}$ calculated via Equation 11. The mean $m_{dr}$ misfit curve across all LR exhibits small spikes near the mode transition points. In practice, the transition from the fundamental to higher modes often occurs smoothly, and the transition frequencies may not correspond precisely to any theoretical mode; this is sometimes referred to as the effective mode. Forcing these points to match a theoretical mode can artificially alter the estimated depth or velocity near a contrast. Therefore, examining the frequency-wise relative misfit of all theoretical dispersion curves is crucial, as high $m_{dr}$ values indicate transition points. During inversion, it is advisable to avoid emphasizing these points, as discussed later in this paper.

Figure 5c shows the inverted $V_s$ from all trials. Several inverted profiles successfully reproduce the velocity contrasts at 3, 8, and 18 m accurately; however, the contrast at 28 m exhibits uncertainty. This behavior is expected, as deeper interfaces are less well constrained by the available frequency band of the simulated dispersion data. In contrast, the fundamental-mode inversion significantly overestimates velocities between roughly 4 and 30 m depth. Finally, Figure 5d compares the best-fit profiles for each predominant-mode LR parameterization with those from the best-fit fundamental-mode inversions. The predominant-mode inversions successfully capture several velocity contrast boundaries, showing good agreement with the true profile, while some deviation occurs at greater depths, whereas the fundamental-mode inversions deviate strongly across the shallow to intermediate depth ranges.

To quantify the accuracy of the inverted $V_s$ profiles, the velocity misfit, $m_{V_s}$ (Equation 10) was computed for all inversion results and is presented in Figure 6. Each vertical bar represents the range of $m_{V_s}$ values across the ten inversion trials performed for each layering ratio. The bottom of each bar corresponds to the lowest (best) $m_{V_s}$ obtained among the ten trials, and the top corresponds to the highest $m_{V_s}$ for that parameterization. The colored solid bars correspond to the predominant-mode inversion results from this study, the gray bars represent the $m_{V_s}$ from the fundamental-mode inversion of the experimental dispersion data extracted from wavefield simulations performed in this study, and the hollow bars denote the benchmark $m_{V_s}$ values reported by Vantassel and Cox (2021) for a fundamental mode target and fundamental mode inversion. Across all LRs (1.5, 2, and 3), the predominant-mode inversion consistently produces much lower $m_{V_s}$ values than the fundamental-mode inversions, confirming that forcing the dispersion data to follow the fundamental mode leads to poorer inversion accuracy.

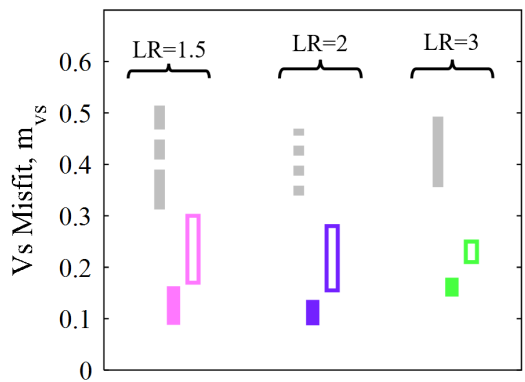

**Figure 6.** Comparison of the $V_s$ misfit values ($m_{V_s}$) for Model g across different LRs: gray bars = fundamental mode inversions performed for this study, colored solid bar = predominant-mode inversions performed for this study, and colored hollow bar = Vantassel and Cox (2021).

### 5.1.2. Ground Model h

Model h is a 6-layer model with $V_s$ boundaries at 2, 5, 8, 16, 21, and 31 m depth, underlain by a half-space. Similar to Model g, this case was designated by Vantassel and Cox (2021) as a high-variance dataset that is challenging to invert accurately. While no single boundary exhibits a large velocity contrast, the cumulative effect of multiple moderate contrasts generates modal osculation in the dispersion response (refer to Figure 3b,e). Figure 7 presents the results of the predominant-mode inversion alongside those obtained from the fundamental-mode inversion of the simulated experimental dispersion data. The target dispersion data and inversion-derived theoretical curves (100 per LR) are shown in Figure 7a. For the predominant-mode inversion (colored curves), the theoretical dispersion curves (i.e., R0 and R1) accurately follow the target dispersion data across most of the frequency band ($m_d$ values range from 0.14-0.28, depending on LR) and successfully capture the transition from the R0 to the R1 mode near 7–8 Hz. Small deviations appear near the transition region, which reflects the inherent difficulty of fitting effective modes in cases where the observed dispersion data do not align precisely with a single theoretical branch. The fundamental-mode inversion (gray curves) follows the overall shape of the target, but poorly fits the dispersion data, producing visibly larger $m_d$ values (0.74-1.48, depending on LR).

The $m_{dr}$ value between target and predominant mode shown in Figure 7b remains low (<5%) across most frequencies for predominant-mode inversion, with isolated peaks near the modal transition. These misfit spikes emphasize the need to carefully evaluate transition frequencies rather than forcing them into specific modes, which risks biasing velocity structure estimates. A strategy for accounting for these transition points is discussed below in regards to Model k.

The suite of inversion-derived $V_s$ profiles (Figure 7c) demonstrates stable reconstruction of the shallow velocity structure for the predominant-mode inversion, with clear recovery of the contrasts at 2, 5, and 8 m across all layering ratios. At intermediate depths (16–21 m), LR = 1.5 and LR = 2 preserve the site layering better, whereas LR = 3 has fewer layers that force averaging in $V_s$ across actual site layer boundaries. The fundamental-mode inverted profiles overestimate $V_s$ consistently below 5 m.

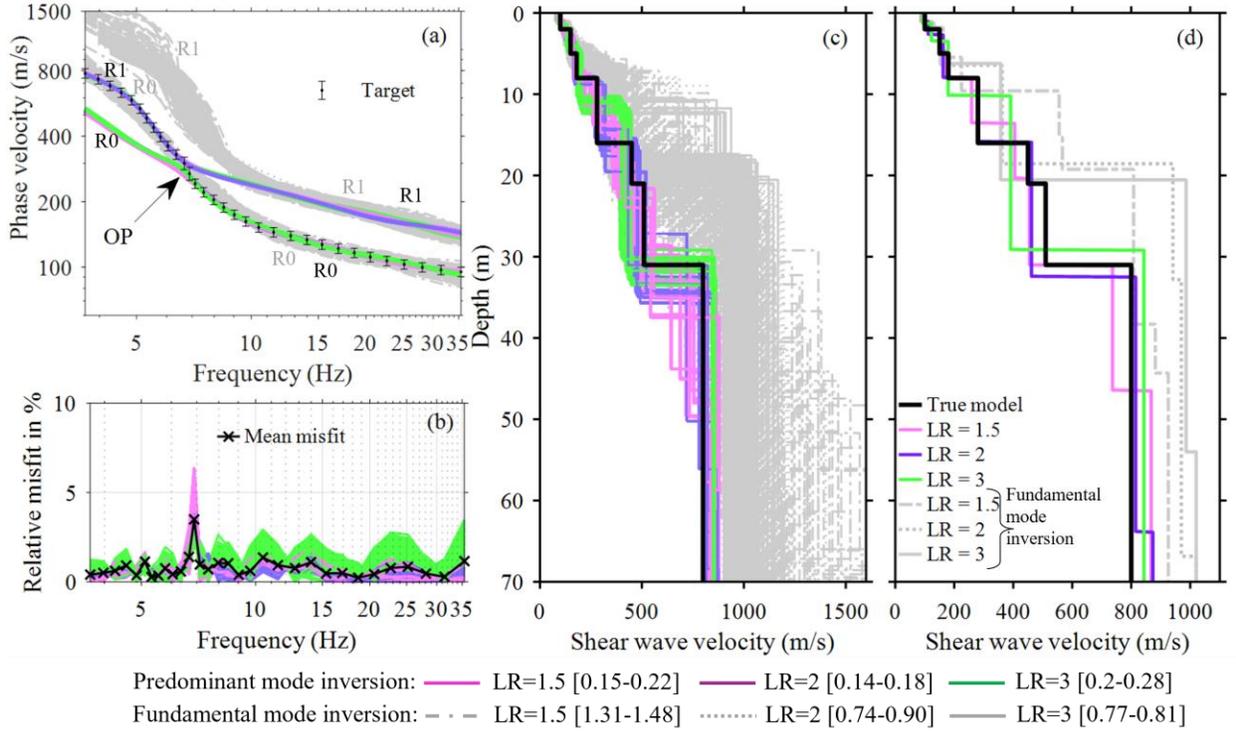

**Figure 7.** Inversion results for Model h: (a) dispersion data (black points with error bars) and inversion-derived theoretical dispersion curves for the ten best-fit solutions from each trial (100 models per LR); colored curves correspond to predominant-mode inversion; gray curves correspond to fundamental-mode inversion; (b) frequency-wise relative misfit between the theoretical dispersion curves and observed dispersion data, with the black line showing the mean misfit trend across all trials; (c) corresponding inversion-derived $V_s$ profiles obtained from the same ten best-fit solutions for each trial (100 models per LR); and (d) best-fitting profiles from each predominant-mode LR inversion compared with the best fundamental-mode inversion results. Misfit values shown in brackets with each LR denote the range in dispersion misfits for the corresponding set of 100 models.

Figure 7d compares the best-fitting $V_s$ profiles for each predominant-mode LR with those from the best fundamental-mode LR inversion. All predominant-mode best-fit $V_s$ profiles agree closely with the true $V_s$ profiles through the upper 20–25 m, with LR = 1.5 and LR = 2 providing the closest match at deeper interfaces. LR = 3 slightly underestimates the contrast near 31 m. In contrast, the fundamental-mode inversion results depart substantially from the true $V_s$ profile at all depths greater than 5 m.

Figure 8 presents the $m_{V_s}$ values for Model h across the three layering ratios, following the same bar-type convention described above for Model g. The predominant-mode inversion yields consistently lower $m_{V_s}$ values across all layering ratios, with best-fit values typically below 0.15 and only small variability between trials. In contrast, the fundamental-mode inversion exhibits substantially higher misfits, generally between 0.30 and 0.55. The benchmark values reported by Vantassel and Cox (2021) are, in general, slightly higher than the predominant-mode results from this study. Overall, the misfit distribution demonstrates that the predominant-mode inversion provides a markedly more accurate reconstruction of the $V_s$ profile for Model h than the fundamental-mode inversion.

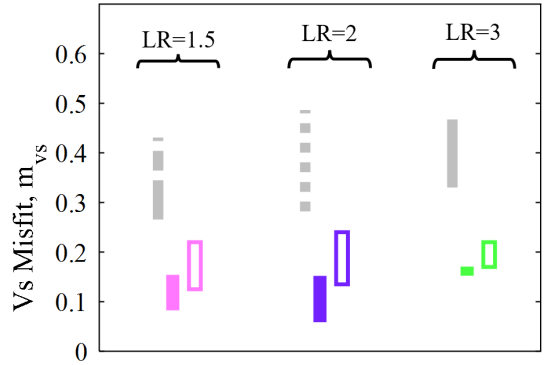

**Figure 8** Comparison of the $V_s$ misfit values ($m_{V_s}$) for Model h across different LRs: gray bar = fundamental mode inversions performed for this study, colored solid bar = predominant-mode inversions performed for this study, and colored hollow bar = Vantassel and Cox (2021).

### 5.1.3. Ground Model k

Model k is a 4-layer model with $V_s$ boundaries at 4, 7, 17, and 42 m depth, underlain by a half-space. The deepest interface exhibits a strong velocity contrast of more than a factor of 2.2, which promotes modal osculation in the dispersion response at frequency around 3.5 Hz, which is a lower frequency compared to the earlier examples (refer to Figure 3). Figure 9a shows the target dispersion data along with the theoretical curves corresponding to the ten best-fitting profiles for each layering ratio. The predominant-mode inversion follows the target dispersion trend well across the entire usable band and successfully captures the modal transition. In the fundamental-mode inversion, R0 aligns with the target better when compared to the results for Models g and h, as indicated by the relatively low $m_d$ values. The frequency-wise $m_{dr}$ plots in Figure 9b remain low for most frequencies in the predominant-mode inversion, with a prominent spike near 3.5–4 Hz, where the modal transition occurs. This spike exceeds 5% relative misfit and marks an effective-mode region in which the observed data do not correspond cleanly to any single theoretical branch. Such peaks highlight transition frequencies that should not be forced to match a particular mode during inversion, as discussed in greater detail below. The full suite of inverted $V_s$ profiles shown in Figure 9c demonstrates that the predominant-mode inversion reconstructs the shallow and intermediate true $V_s$ profile reliably. The contrasts at 4, 7, and 17 m are captured well across all layering ratios, and the deeper contrast at 42 m is recovered with some variability. The fundamental-mode inversion, in contrast, tends to overestimate velocities below approximately 30 m because it cannot represent the higher-mode contribution in the target dispersion data. Similarly, the best-fit profiles of the predominant mode inversion track the true $V_s$ profile well through the upper ~25 m, with noticeable smearing of the deep contrast (Figure 9d). This smearing is attributed to modal transition effects; to examine this, dispersion data with frequencies near the transition points, where $m_{dr}$ values are high (> 5%), were excluded (red dispersion data points in Figure 10a) and the inversion was re-run. Note that a 5% $m_{dr}$ threshold derived from the synthetic analysis was used to identify and exclude dispersion data within the transition-band frequencies. While excluding those dispersion data points inevitably removes information in that frequency band, and thus increases local uncertainty, it avoids forcing a misleading fit that would distort velocities near the contrast.

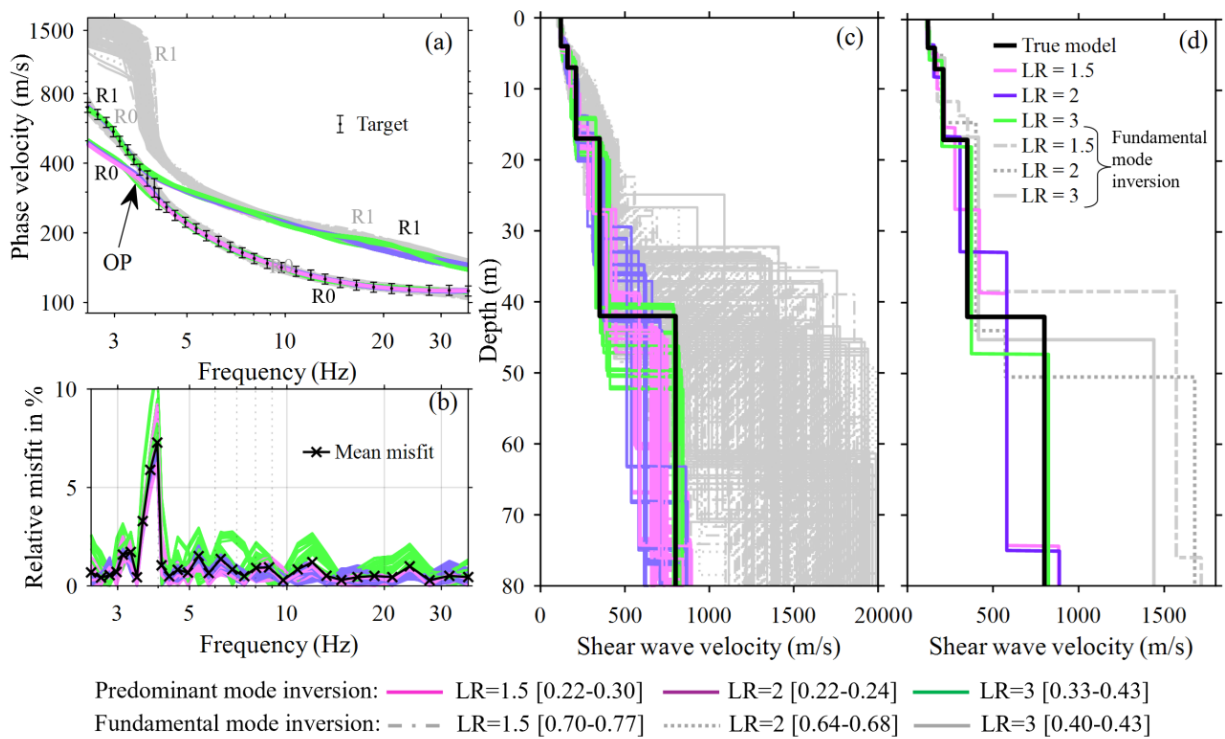

**Figure 9.** Inversion results for Model k: (a) dispersion data (black points with error bars) and inversion-derived theoretical dispersion curves for the ten best-fit solutions from each trial (100 models per LR); colored curves correspond to predominant-mode inversion; gray curves correspond to fundamental-mode inversion; (b) frequency-wise relative misfit between the theoretical dispersion curves and observed dispersion data, with the black line showing the mean misfit trend across all trials ; (c) corresponding inversion-derived $V_s$ profiles obtained from the same ten best-fit solutions for each trial (100 models per LR); and (d) best-fitting profiles from each predominant-mode LR inversion compared with the best fundamental-mode inversion results. Misfit values shown in brackets with each LR denote the range in dispersion misfits for the corresponding set of 100 models.

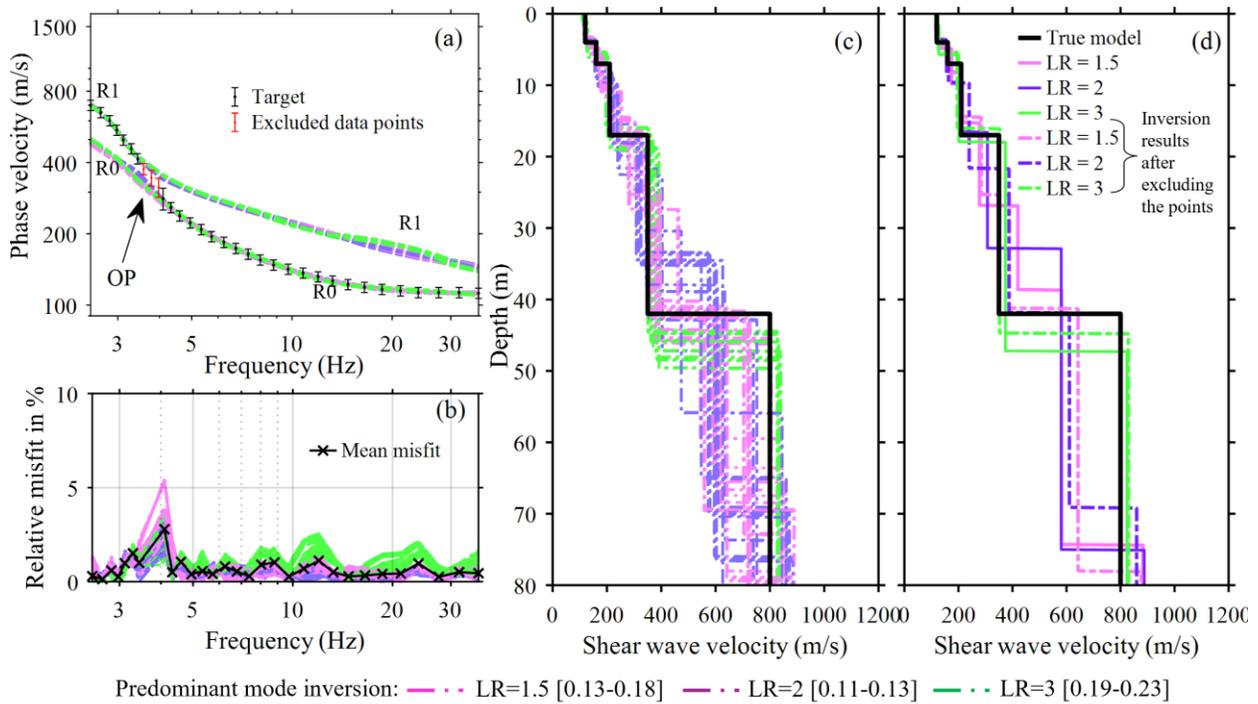

**Figure 10.** Inversion results for Model k after excluding frequency points around the modal transition: (a) (a) dispersion data (black points with error bars, excluded points in red bars) and inversion-derived theoretical dispersion curves for the ten best-fit solutions from each trial (100 models per LR); (b) frequency-wise relative misfit between the theoretical dispersion curves and observed dispersion data, with the black line showing the mean misfit trend across all trials ; (c) corresponding inversion-derived $V_s$ profiles obtained from the same ten best-fit solutions for each trial (100 models per LR); and (d) best-fitting profiles from each predominant-mode LR inversion compared with the best profiles before excluding the points (solid lines), i.e.,in Figure 9d . Misfit values shown in brackets with each LR denote the range in dispersion misfits for the corresponding set of 100 models.

Figure 10 shows the results from re-running the inversions after excluding dispersion data with $m_{dr}$ values great than 5% around the modal transition OP frequency (~3.5 Hz). As shown in Figure 10a, the theoretical curves corresponding to the best-fitting models more closely follow the target dispersion trend across the remaining frequencies, without the mismatch previously observed near the OP in Figure 9a. The corresponding $m_{dr}$ values (Figure 10b) in the modal transition frequency range also exhibit a significant reduction relative to the large spike shown in Figure 9b, confirming that excluding the ambiguous modal transition points prevents the inversion from attempting to satisfy conflicting modal behavior. The full suite of inverted $V_s$ profiles (Figure 10c) demonstrates a tighter clustering of the predominant-mode $V_s$ profiles at depth, particularly around the 42 m $V_s$ contrast, where the original predominant-mode inversion $V_s$ profiles showed the greatest variability. Although some smoothing remains inevitable due to the loss of phase-velocity information at the excluded frequencies, the predominant-mode $V_s$ profiles now exhibit improved consistency with the true $V_s$ profile and have reduced spread. This stabilizing effect is also reflected in the best-fitting predominant-mode $V_s$ profiles (Figure 10d), which show closer agreement with the true model across the full depth range. The deep velocity contrast is more sharply defined than in the original best-fitting inversion results (shown in solid line), and the upper-layer structure remains unchanged, confirming that the exclusion of dispersion data near modal transition OP frequency does not compromise shallower resolution. However, LR = 1.5 and 2 still show some underestimation of $V_s$ for the half-space, while LR = 3 represents the half-space velocity well.

Figure 11 presents the corresponding shear-wave velocity misfit values $m_{V_s}$ for Model k across the three layering ratios, using the similar bar-type convention defined earlier (gray = fundamental-mode inversion, colored solid = predominant-mode inversion, colored dashed = predominant-mode inversion after exclusion of transition frequencies, hollow = values reported by Vantassel and Cox, 2021). For all LR, the predominant-mode inversion produces lower $m_{V_s}$ values than the fundamental-mode inversion, consistent with the behavior observed in the dispersion and velocity-profile comparisons. After excluding the higher relative dispersion misfit transition-zone frequencies and re-running the inversions, the $m_{V_s}$ values decrease further for LR = 1.5 and LR = 2, reflecting the improved recovery of the deeper contrast shown in Figure 10. In contrast, LR = 3 shows a slight increase in misfit variability, attributable to its coarser parameterization and reduced sensitivity at depth. The benchmark misfits reported by Vantassel and Cox (2021) are generally similar to, or slightly greater than those obtained with the predominant-mode inversion results. Overall, the misfit distributions confirm that removing poorly constrained transition-zone frequencies stabilizes the inversion and enhances the accuracy of the recovered $V_s$ structure for Model k.

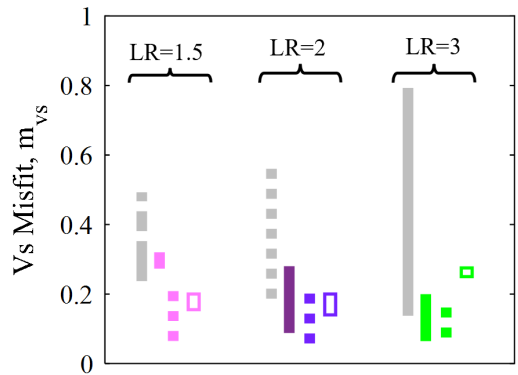

**Figure 11.** Comparison of the $V_s$ misfit values ($m_{V_s}$) for Model k across different LRs: gray = fundamental-mode inversion performed for this study, colored solid = predominant-mode inversion performed for this study, colored dashed = predominant-mode inversion after exclusion of transition frequencies, and colored hollow = values reported by Vantassel and Cox (2021).

These results confirm that the predominant-mode strategy, combined with careful treatment of experimental dispersion data near modal-transition frequencies, yields stable and reliable inversions. By avoiding biased fitting at transition frequency bands, the predominant-mode approach minimizes artificial distortion of the velocity profile and produces solutions that are generally more accurate than those obtained in prior studies.

## 6.1. Field Example

To evaluate the effectiveness of the proposed method when applied to field data, the predominant-mode inversion was applied to a surface wave dataset collected near the I-15 Downhole Array (I15DA) site in Salt Lake City, Utah. Both the surface wave dataset and the invasive borehole data were obtained from Jackson (2024). Active-source MASW testing was performed using a linear array of 24 geophones spaced at 2 m, with shot offsets of 5, 10, and 20 m. Passive data were acquired using three separate MAM arrays, each comprised of 10 broadband seismometers. One circular MAM array had a diameter of 60 m (C 60) and two L-shaped arrays had maximum leg lengths of 300 m (L 300) and 1000 m (L 1000), respectively. The dispersion data were processed following the workflow and open-source tool *SWProcess* developed by Vantassel and Cox (2022). Additional details of the acquisition and processing methodology are discussed in Jackson (2024) and the dataset is published open-access in Cox et al. (2025). The dispersion data with ± one standard deviation bounds obtained from the combined MASW and MAM processing performed by Jackson (2024) was used as the target data for the predominant-mode inversions performed herein. The target dispersion data is shown in Figure 12a. Notably, the processed dispersion data did not exhibit any visible evidence of higher-mode branches at low frequencies and was originally interpreted by Jackson (2024) to consist only of fundamental mode data.

The predominant-mode inversion was carried out using four different layering ratios (LR = 1.5, 2, 3, and 5). The limits of maximum and minimum layer thickness were set based on the recommendations of Cox and Teague (2016) and the minimum and maximum $V_s$ values were constrained between 100 and 3500 m/s. First, the inversion was performed using the full dispersion data reported by Jackson (2024). However, when plotting the $m_{dr}$ values following initial inversions, three data points were observed to exhibit high mean $m_{dr}$ values of approximately 6–9% within the frequency range of 1.36–1.55 Hz, with several individual LR theoretical curves exceeding 10% (not shown in Figure 12). To reduce bias from effective-mode transitions, dispersion data points with mean $m_{dr}$ values exceeding 5% were excluded, following the procedure demonstrated for synthetic Model k. After excluding the modal transition frequencies, the inversion was re-run, and the resulting profiles are presented in Figure 12. For each LR, five independent trials were conducted, and the single best-fitting profile from each LR was selected for plotting. These are shown together with the invasive $V_s$ measurements obtained from downhole PS-logging at the site to a depth of 116 m (Youd & Briggs, 2003). The theoretical dispersion curves (Figure 12a) fit the experimental data well and indicate that the predominant-mode inversion captured the target dispersion data well across the usable frequency range, while the frequency-wise $m_{dr}$ value (Figure 12b) confirms that most of the inversion error is concentrated at a few isolated frequencies. The inversion-derived $V_s$ profiles demonstrate good agreement with the PS-logging data in the upper ~100 m, particularly for LR = 1.5 and 2, which capture the shallow velocity structure with reasonable reliability (Figure 12c). In Figure 12d, the profiles are extended to a depth of 1,000 m, which is equal to the aperture of the largest MAM array and is less than ½ the maximum resolved dispersion wavelength.

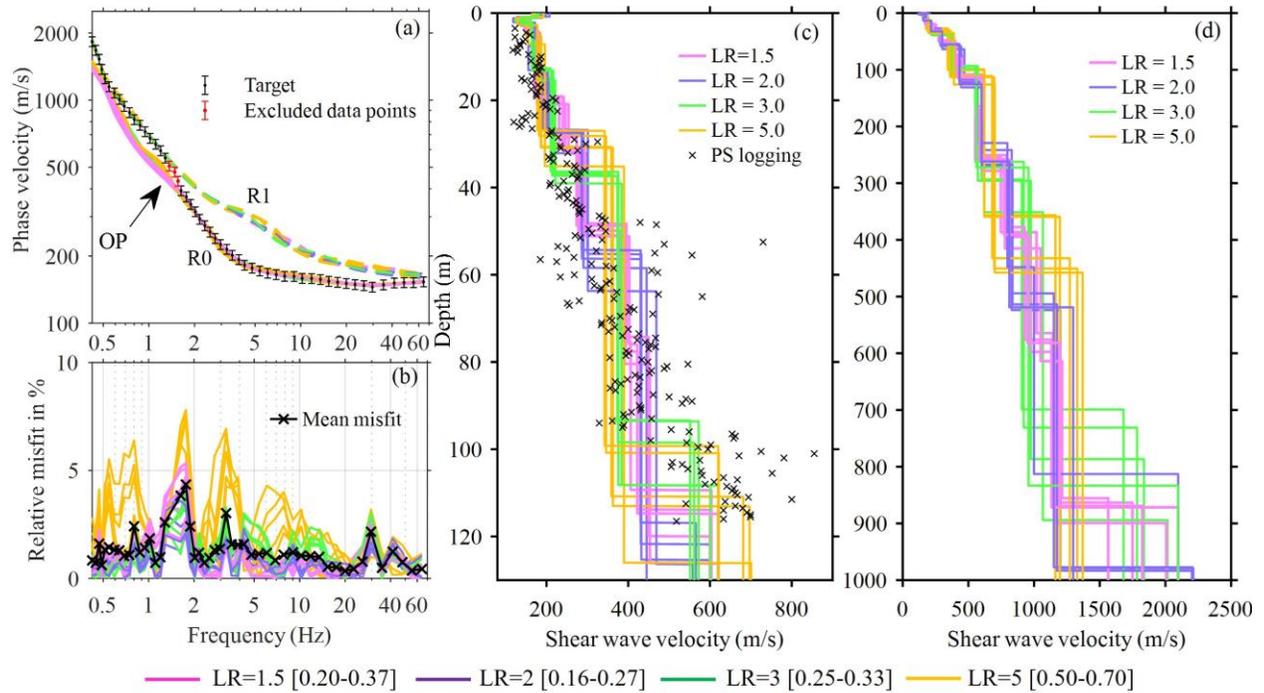

**Figure 12.** Inversion results for the I15 downhole array site: (a) observed dispersion data (black with error bars and excluded points in red) and inversion-derived theoretical dispersion curves for the lowest-misfit solutions from each trial (5 models per LR); (b) frequency-wise relative misfit between the theoretical dispersion curves and observed dispersion data, with the black line showing the mean misfit trend across all trials; (c) corresponding inversion-derived $V_s$ profiles from the best-fit solutions for each trial (5 models per LR), compared with invasive PS-logging data (Youd & Briggs, 2003); and (d) extended best-fitting $V_s$ profiles from each LR plotted to greater depth. Misfit values shown in brackets with each LR denote the range in dispersion misfits for the corresponding set of 5 models.

To further validate the $V_s$ profiles obtained from the predominant-mode inversion, the Rayleigh-wave ellipticity angle was extracted at each frequency from the 3-component MAM data using the wavefield decomposition method (Wathelet et al., 2018). The ellipticity measurements shown in Figure 13 display a clear shift in particle-motion direction from prograde (positive ellipticity) to retrograde (negative ellipticity) at approximately 1.5 Hz. Changes in Rayleigh wave ellipticity angle often occur at frequencies where Rayleigh wave mode transitions occur in dispersion data. This frequency closely matches the autodetected OP identified by the predominant-mode inversions, as shown in Figure 12a, providing independent confirmation of the mode interpretation. Such a change in ellipticity sign is commonly associated with strong impedance contrasts in the subsurface, which modify the horizontal-to-vertical displacement ratio (Bergamo et al., 2023). Although Rayleigh-wave ellipticity sign changes can also coincide with peaks in the Rayleigh-wave ellipticity amplitude, and subsequently with the peak in the horizontal-to-vertical spectral ratio (H/V), no corresponding H/V peak was observed at this site near the 1.5 Hz band (Jackson, 2024). Therefore, the observed sign reversal is most likely attributable to a transition between modes rather than a resonance or amplification effect.

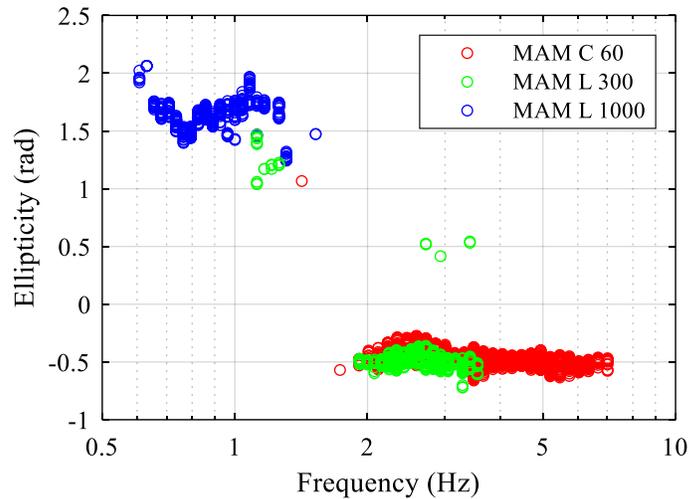

**Figure 13.** Ellipticity angle extracted from passive MAM data for three array configurations (C60, L300, and L1000), showing the frequency-dependent particle-motion polarity.

### 6.1.1. Comparing Transfer Functions

To further assess the accuracy of the predominant-mode inverted $V_s$ profiles, theoretical transfer functions (TTF) derived from the inverted $V_s$ profiles are compared with empirical transfer functions (ETF) obtained from aftershock ground motions recorded by the I15 Downhole Array (Dawadi et al., 2025; Jackson & Cox, 2026). The I15DA consists of five three-component sensors installed at the surface and at depths of 7.6, 18.3, 48.8, and 119.8 m. Jackson (2024) computed the ETF between the surface and the deepest sensor at 119.8 m using a suite of aftershock earthquake recordings at small strain levels. Figure 15 compares their median ETF with the 1D TTFs calculated from the inverted predominant-mode $V_s$ profiles determined in this study. The theoretical shear wave TTFs were calculated assuming a linear-viscoelastic horizontally layered medium overlaying elastic bedrock (Kramer, 1996). "Within" boundary conditions were applied at the depth corresponding to the deepest sensor at 120 m (Teague et al., 2018). The damping ratios for the soil layers were assigned using the Darendeli (2001) model, and a constant damping ratio of 0.5% was used for the rock. The ETF plotted here represents the lognormal median of both the north–south and east–west horizontal components; the detailed processing and quality-control procedures are described in Jackson (2024). The ETF clearly exhibits four distinct resonance peaks at approximately 0.8, 1.85, 2.9, and 4.17 Hz. These ETF resonance peaks can serve as benchmarks for validating the inverted $V_s$ profiles, as TTFs calculated from accurate $V_s$ profiles should reproduce these resonance peaks reasonably well. The TTFs were calculated from the 20 best-fit $V_s$ profiles obtained from each of the five inversion trials using LR = 1.5, 2, 3, and 5. The TTFs show good consistency with the ETF resonant frequencies, particularly at the first two (lowest frequency) peaks, indicating that the predominant-mode inversion successfully captured the key impedance contrasts in the subsurface. Note that we are not expecting amplitude matches between the TTFs and the ETF, as the 1D TTFs cannot correctly capture apparent damping/attenuation caused by non-1D site effects, like non-vertical wave incidence and lateral variability in soil layering that influence the actual recorded ground motions (Dawadi et al., 2025). Nonetheless, the suit of predominant-mode inversion-derived $V_s$ profiles produce TTFs that align closely with the frequencies amplified by empirical ground motions recorded at the site. This agreement provides independent validation of the predominant-mode inversion results, beyond the dispersion-data fitting, and highlights the method's ability to recover $V_s$ profiles that are consistent with both surface wave data and observed earthquake ground motions.

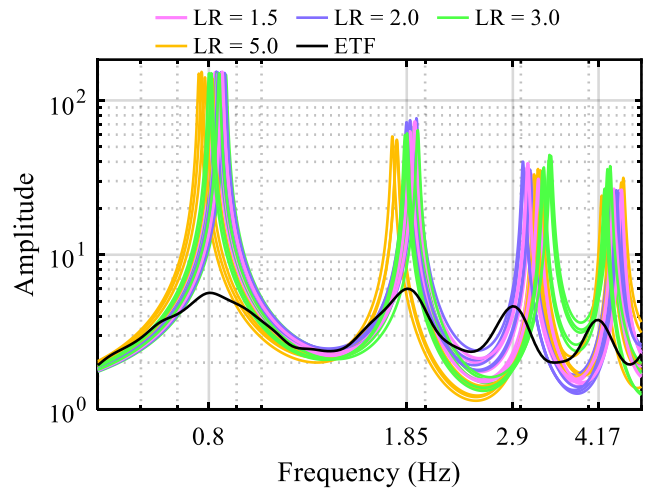

**Figure 14.** Comparison between ETF (black line) and TTF (colored curves) computed from the best-fitting inversion-derived profiles for different LRs.

## 7.1. Conclusions

Surface wave methods are among the most effective non-invasive tools for geotechnical site characterization. Their ability to derive $V_s$ profiles over a wide depth range, particularly when active and passive surface wave measurements like MASW and MAM are combined, makes them invaluable for both shallow and deep site investigations. A key challenge, however, lies in the inversion of surface wave dispersion data collected at sites with strong subsurface velocity contrasts. When velocity contrasts greater than approximately 2.0 are present, surface wave energy can gradually transition from the fundamental mode at high frequencies to a higher mode at low frequencies, a phenomenon commonly referred to *as* mode osculation. The transition of modal energy due to strong impedance contrast is often smooth and continuous in the calculated dispersion data due to a lack of spatial resolution in the experimental acquisition sensor arrays. As a result, misidentifying a higher surface wave dispersion mode as the fundamental mode frequently leads to systematic overestimation of $V_s$ in deeper layers and/or misinterpretation of bedrock depth during inversion. Although effective mode (or superimposed mode) inversion algorithms can theoretically account for such transitions, effective mode computations require precise knowledge of the source–receiver configuration used to collect the dispersion data. Consequently, effective-mode formulations work well only for MASW records collected with an individual shot. In practice, however, target dispersion data used for inversion are often constructed by combining dispersion data extracted from multiple MASW shots with varying offsets and shot directions with dispersion data extracted from several 2D MAM array recordings for which source locations are unknown. This combination of active MASW and passive MAM dispersion datasets makes the implementation of an effective mode inversion untenable.

To address the mode osculation problem at low-frequencies, this study introduced a predominant-mode inversion framework that does not require manual mode identifications made by the inversion analyst or explicit information about the source and receiver geometries used to collect the dispersion data. Rather, the predominant-mode framework automatically identifies, at each frequency, the Rayleigh-wave mode with the maximum vertical amplitude at the free surface, enabling inversions to proceed without requiring prior knowledge of mode order or assumptions about which mode branch dominates the recorded wavefield. The predominant-mode forward modelling was carried out using the thin layer method. To evaluate its performance, the predominant-mode forward problem was incorporated into a particle swarm optimization algorithm and used to invert dispersion data from three synthetic models representing different levels of complexity, each displaying smooth low-frequency transitions of modal energy caused by strong impedance contrasts. The inversion targets were generated by simulating active and passive surface-wave tests, ensuring realistic dispersion data. Inversions were then performed across multiple layering parameterizations to account for inversion layering uncertainty. In each synthetic study case, the predominant-mode method successfully captured major velocity contrasts, respective depths, and the associated osculation point in the dispersion data with good accuracy. However, traditional inversions based on a fundamental mode assumption consistently overestimated $V_s$ and mis-represented layer boundaries. The predominant-mode inversion was further evaluated using a field dataset of active MASW and passive MAM dispersion data collected near the I15 Downhole Array Site. The inverted $V_s$ profiles were first compared against downhole PS-logging data available at the site to a depth of 116 m, wherein the inverted $V_s$ profiles were shown to agree well with those from invasive $V_s$ measurements. Additionally, the predominant-mode $V_s$ profiles were further validated by comparing the ETF derived from recorded earthquake ground motions at the site with the TTFs of the best-fitting inverted $V_s$ profiles. The ETF-to-TTF comparisons showed good agreement between the measured and predicted resonant frequencies amplified by the subsurface, confirming that the $V_s$ profiles from the predominant-mode inversion captured the key impedance contrasts controlling seismic site response. The transition of modal energy in the dispersion data was further confirmed by examining Rayleigh-wave ellipticity angles extracted from the passive-wavefield MAM measurements. The osculation point identified in the predominant-mode inversion was found to coincide with the frequency range where the Rayleigh-wave particle motion switched from prograde to retrograde, consistent with the expected ellipticity sign change. Overall, the predominant-mode inversion framework offers a robust alternative to conventional inversion approaches by automatically mitigating mode misidentification and reducing velocity overestimation at sites with strong impedance contrasts, particularly when dispersion data is extracted from multiple MASW shot locations and/or when MASW and MAM dispersion data are combined.

## CRediT authorship contribution statement

**Mrinal Bhaumik:** Writing -original draft, Visualization, Methodology, Validation, Coding. **Brady Cox:** Writing-review & editing, Supervision, Conceptualization, Methodology, Validation.

## Declaration of competing interest

The authors declare that they have no known competing financial interests or personal relationships that could have appeared to influence the work reported in this paper.

# Data availability

The synthetic models used in this study are available through the NHERI DesignSafe Data Repository at https://doi.org/10.17603/ds2-cpmr-v194 . The field data set is also accessible through DesignSafe data repository at https://doi.org/10.17603/ds2-xt4m-xa77. The MATLAB code implementing the predominant-mode inversion framework will be made publicly available upon publication.

# Acknowledgements

The authors gratefully acknowledge the use of the field dataset originally processed and documented by Tyler Jackson as part of his master's research at Utah State University. The data were accessed through the NHERI DesignSafe-CI Data Repository, whose archival resources were essential for this study. The authors also thank the Department of Civil and Environmental Engineering at Utah State University for providing additional computational resources. Any opinions, findings, and conclusions expressed in this paper are those of the authors and do not necessarily reflect the views of any affiliated institution.